\documentclass[aip,jap,reprint,showpacs,showkeys,graphicx]{revtex4-1}

\usepackage{graphicx}
\newcommand{\degC}{\ensuremath{^{\circ}\text{C }}}
\begin{document}


\title{Full field electron spectromicroscopy applied to ferroelectric materials} 

\author{N. Barrett}
\email{nick.barrett@cea.fr}
\author{J. E. Rault}
\author{J. L. Wang}
\author{C. Mathieu}
\affiliation{IRAMIS/SPCSI/LENSIS, F-91191 Gif-sur-Yvette, France}
\author{A. Locatelli}
\author{T. O. Mentes}
\affiliation{Sincrotrone Trieste S.C.p.A., Basovizza, Trieste 34149, Italy}
\author{M. A. Ni\~{n}o}
\affiliation{Instituto Madrileño de Estudios Avanzados en Nanociencia (IMDEA-Nanociencia), Campus de Cantoblanco, Cantoblanco s/n 28049 Madrid, Spain}
\author{S. Fusil}
\affiliation{Universit\'{e} d'Evry-Val d'Essonne, Boulevard Fran\c cois Mitterrand, 91025 Evry cedex, France}
\author{M. Bibes}
\author{A. Barth\'{e}l\'{e}my}
\author{D. Sando}
\affiliation{Unit\'{e} Mixte de Physique CNRS/Thales, 1 Av. Augustin Fresnel, Campus de l'Ecole Polytechnique, 91767 Palaiseau, France}
\author{W. Ren}
\affiliation{Physics Department, University of Arkansas, Fayetteville, Arkansas 72701, USA}
\affiliation{Department of Physics, Shanghai University, 99 Shangda Road, Shanghai 200444, China}
\author{S. Prosandeev}
\author{L. Bellaiche}
\affiliation{Physics Department, University of Arkansas, Fayetteville, Arkansas 72701, USA}
\author{B. Vilquin}
\affiliation{Universit\'{e} de Lyon, Ecole Centrale de Lyon, Institut des Nanotechnologies de Lyon, F- 69134 Ecully cedex, France}
\author{A. Petraru}
\affiliation{Nanoelektronik, Technische Fakult\"{a}t, Christian-Albrechts-Universit\"{a}t zu Kiel, D-24143 Kiel, Germany}
\author{I. P. Krug}
\author{C. M. Schneider}
\affiliation{Peter Gr\"{u}nberg Institute (PGI-6) and JARA-FIT, Research Center J\"{u}lich, D-52425
J\"{u}lich, Germany}




\date{\today}
 
\begin{abstract}
The application of PhotoEmission Electron Microscopy (PEEM) and Low Energy Electron Microscopy (LEEM) techniques to the study of the electronic and chemical structure of ferroelectric materials is reviewed. Electron optics in both techniques gives spatial resolution of a few tens of nanometres. PEEM images photoelectrons whereas LEEM images reflected and elastically backscattered electrons. Both PEEM and LEEM can be used in direct and reciprocal space imaging. Together, they provide access to surface charge, work function, topography, chemical mapping, surface crystallinity and band structure. Examples of applications for the study of ferroelectric thin films and single crystals are presented. 
\end{abstract}

\maketitle 

\section{INTRODUCTION}
The aim of this overview is to introduce and illustrate the use of full field spectromicroscopy techniques such as PhotoEmission Electron Microscopy (PEEM) and Low Energy Electron Microscopy (LEEM) to study the electronic and chemical structure of ferroelectric (FE) materials. Particular emphasis will be given to the study of thin films and their electrical boundary conditions. The term spectromicroscopy is used to stress that these techniques yield spectroscopic and microscopic information simultaneously. This is to be distinguished from microspectroscopy  which, in order to obtain spatial information over similar fields of view, relies on scanning a microfocussed beam, whose size defines the spatial resolution, over the surface. This article is not intended to be an exhaustive review, nor is it aimed at PEEM or LEEM specialists. Rather, we hope to give scientists working on ferroic systems a better idea of the variety of information which may be obtained using spectromicroscopy on domain structured materials.

In the last two decades, three main breakthroughs have led to intense research into ferroelectrics and more generally ferroic materials. On a theoretical level,  Cohen showed that the ferroelectric stability depends on a balance between long and short range forces. \cite{Cohen1992} Vanderbilt and King-Smith developed the modern theory of polarization using a Berry phase approach which allowed the description of the change in polarization as a displacement of the centre of charge of the Wannier functions.  \cite{King-Smith1993} Nowadays, approaches using effective Hamiltonians and in some cases kinetic Monte-Carlo methods are starting to bridge the gap between the 100 atom systems of first principles calculations and experimentally observed domains sizes.\cite{Prosandeev2012}
In experiment the widespread use of Piezo response Force Microscopy (PFM) in air, controlled atmosphere and now increasingly under vacuum, allows nanoscale domain writing and measurement of both lateral and perpendicular piezo coefficients. \cite{Kalinin2010, Infante2010}
Finally, on a technological level, advances in epitaxial thin film growth have allowed the production of high quality, single crystal epitaxial thin films, suitable for ferroelectric based devices. Varying the substrate lattice parameter allows strain engineering which can significantly change values of remanent polarization, coercive field or the Curie temperature. \cite{Schlom2007, Choi2004} In fundamental research, two deposition methods are favoured, Molecular Beam Epitaxy (MBE) for near perfect layer by layer growth and Pulsed Laser-assisted Deposition (PLD) for its flexibility and control of oxygen stoichiometry.

The new physics emerging from two-dimensional films in the limit of a few unit cells has a host of exciting applications. Understanding the ferroelectric properties of such engineered thin film systems requires taking into account not only the material but also its interfaces with electrodes, substrates or atmosphere; in other words, the electrical boundary conditions. \cite{Dawber2005} In the case of a thin film these can even determine the ferroelectric polarization stability. For example, Junquera and Ghosez showed that the depolarizing field could place a lower limit on the film thickness capable of supporting a stable polarization. \cite{Junquera2003}
The question of the interface is also a key issue in realizing multiferroic heterostructures, demonstrating magnetoelectric coupling allowing electrical control of the magnetization or, inversely, magnetic control of the polarization. \cite{Ramesh2007} Hybridization between filled d orbitals responsible for magnetization and empty d orbitals in the ferroelectric oxide may be one path to such coupling. Several coupling mechanisms have been identified and these can be quite complex. For example, recent results suggest that even the charge ordering of a magnetic layer can be modulated by the polarization state of an adjacent ferroelectric. \cite{Vaz2010} Since these are collective electron phenomena, high resolution spectroscopy is necessary to untangle the electronic structure responsible for such phenomena.
Thus, experimental tools which can contribute to a full understanding of the surface and near surface electronic and chemical structure are desirable, if possible with spatial resolution suitable for investigating individual domains or even phenomena localized at domain walls.

Near-field techniques such as PFM or scanning surface potential microscopy (SSPM) occupy a prime position in the investigation of ferroelectric domain structure. Like all techniques they require careful use, and attention must be paid to the role of adsorbates and to interactions between the tip and the surface in contact mode. However, they cannot probe the full electron structure and chemistry of ferroelectric surfaces. In view of this, and given the increasing number of fields of applications of engineered ferroelectric systems (for example in promoting chemisorption \cite{Li2008}, in catalysis \cite{Giocondi2001} or for photovoltaics \cite{Seidel2011}), complementary experimental techniques are required to understand these complex systems. Early examples of PEEM applied to ferroelectrics studied adsorbate induced variations in the electron affinity. \cite{Yang2004, Lev2005}

Full field electron microscopy, covering PEEM, LEEM and Mirror Electron Microscopy (MEM), has recently started to provide interesting data on FE surfaces. As a result, it appears opportune to review the physics which can be revealed by the use of these techniques. It is important to note that PEEM may also be performed in X-Ray absorption mode (XAS) by measuring the electron yield from the sample as a function of the incident photon energy. Mainly because of the high intensity this technique is better established and has already been used to study ferroic systems. The Scholl group in Berkeley have pioneered the application of X-ray absorption PEEM for the study of ferroics. \cite{Zhao2006} They demonstrated the use of X-ray magnetic circular dichro\"{i}sm (XMCD) PEEM as a powerful method to image magnetic domain ordering. \cite{Heron2011} The present review concentrates on LEEM and on photoemission based PEEM to probe the surface electronic structure, particularly relevant for the study of ferroelectrics.

In section II the principles behind PEEM and LEEM will be presented with an emphasis on the photoemission process underlying PEEM based methods. The main part of this review consists of the different case studies in section III. The five examples of studies of ferroelectric surfaces which can be carried out using PEEM and LEEM are not meant to be exhaustive, but rather give an idea of the rich information obtainable using spectromicroscopy. The first example illustrates the relationship between the work function as measured in PEEM and the surface polarization charge. The following two studies look at screening by surface carbon and photogenerated electron-hole pairs. The fourth example illustrates how PEEM and MEM-LEEM can be used to estimate the polarization of ultra-thin, tunneling regime films. The last example demonstrates the measurement of the full 2D band structure parallel to the surface of a single, micron sized in-plane ferroelectric domain.  Indeed, there is much more to be done, both in terms of quantification but also in fully exploiting the spectral information available. Section IV discusses both the advantages and present limitations/drawbacks of LEEM and PEEM for studies of FE systems. A short outlook completes the overview in section V.

\section{SPECTROMICROSCOPY}

\begin{figure*}[ht]
\includegraphics[width=14cm,clip,bb=0 0 624 477]{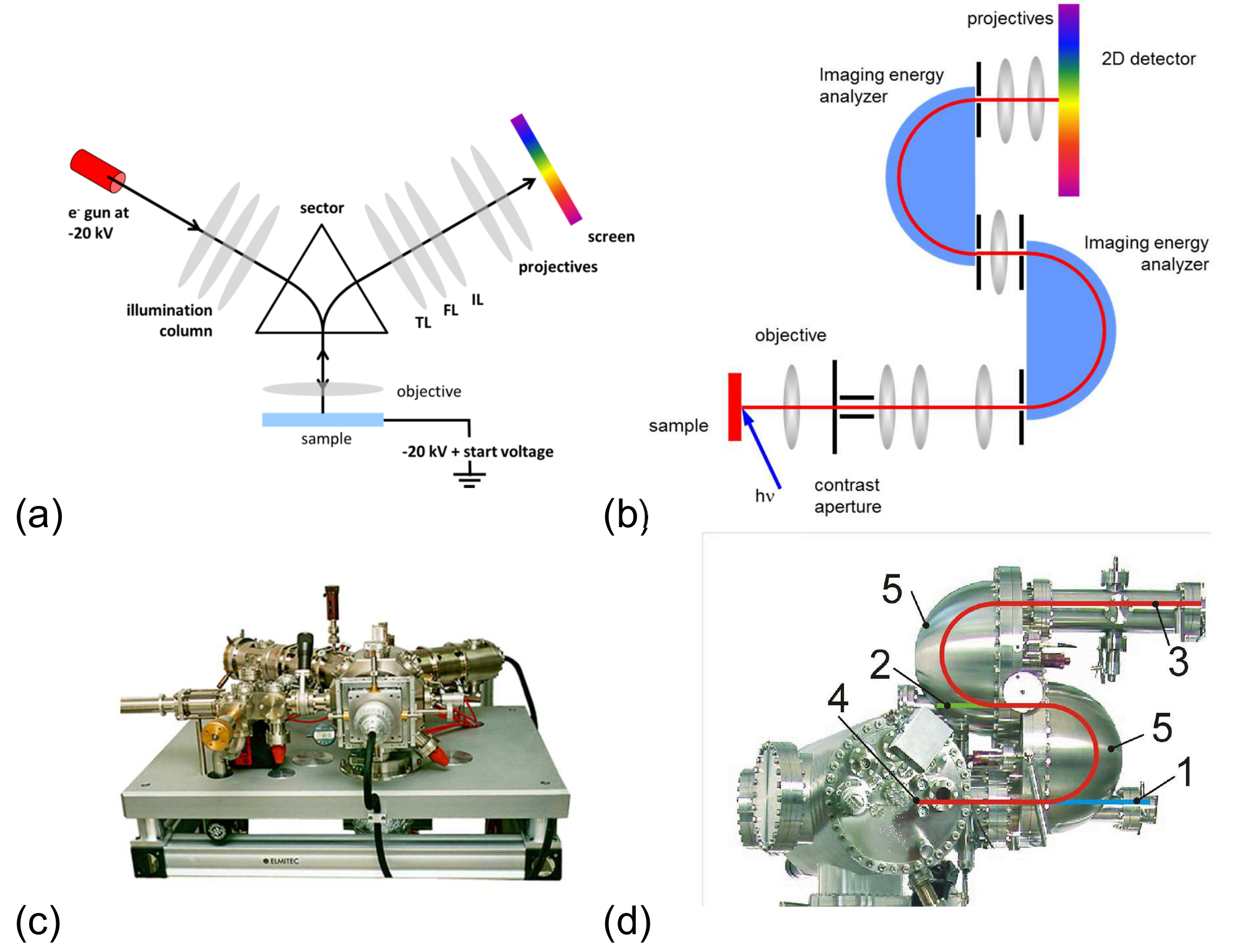}
\caption{(color online) (a) Layout of a LEEM/PEEM instrument (LEEM III, Elmitec GmbH) with magnetic lenses. The sector field separates the incoming electron beam from the reflected or backscattered electrons. The sample bias adjusts the electron start voltage. TL, FL and IL are the transfer, field and intermediate lens of the imaging column. In PEEM mode, an energy analyzer is used just before the last projective lenses.
(b) Layout of an energy-filtered PEEM with electrostatic lenses (NanoESCA, Omicron Nanotechnology GmbH). The double hemispherical analyzer and the low electron energy in the PEEM column optimizes transmission and gives a lateral resolution independent of the spectroscopic resolution.
(c,d) Photographs of typical instruments represented in (a) and (b). In (d) 1, 2 and 3 are the positions for direct PEEM, small area spectroscopy and spectroscopic imaging, 4 and 5 are the sample position and the double hemispherical analyzer.
}
\label{fig:Instruments}
\end{figure*}

PEEM and LEEM are two distinct, surface imaging techniques; however, they are also extremely complementary. Not only are the electron optics of the imaging column often very similar, if not identical, the spatial resolution is often comparable. However, in the case of PEEM the fundamental process is the photoelectric effect, i.e. photon in and electron out, whereas in the case of LEEM the probe is a collimated electron beam and the elastically reflected or backscattered electrons are detected. LEEM is therefore an electron in and electron out process. The information contained in PEEM and LEEM can be very different and the combined use of both techniques provides extremely rich information.

Both systems have been used in the examples given in this review. Recent reviews provide more extensive technical details. \cite{Escher2010, Schneider2012, Locatelli2008a} Here, we just describe the main characteristics of each design.
The first PEEM experiment was reported by a German physicist, Br\"{u}che, in the 1930s. \cite{Bruche1933} However, the real rise of LEEM and PEEM started with the use of more sophisticated, low energy electron lenses in ultra-high vacuum and the pioneering work by Telieps and Bauer in the 1980s. \cite{Telieps1985} The second qualitative step was the advent of dedicated synchrotron radiation sources, capable of delivering a high brilliance photon beam onto the sample, making sub-micron scale PEEM imaging possible in a reasonable amount of time.

First, we outline the two main designs of low energy electron optics. Then we present the different imaging techniques used for the case studies presented in this review.

\subsection{Electron optics}
Two basic design concepts exist in low energy electron microscopes. In both cases there is a high extractor field, typically 12 to 24 kV, between the sample and the objective lens. Thanks to this field, high electron take-off angles from the sample surface can be collected and imaged through the electron optics. In the first design (Fig.~\ref{fig:Instruments}(a)), a very high voltage (20 kV) is applied to the sample, thus the electrons have a high kinetic energy in the electron optics imaging column, which is composed of magnetic lenses. The main manufacturers are Elmitec and SPECS GmbH. In the second (Fig.~\ref{fig:Instruments}(b)), the sample is biased close to ground potential and after extraction by a high voltage the electrons are decelerated in the electrostatic lens system. The main manufacturer is Omicron Nanotechnology.

\subsubsection{Magnetic lenses}
The magnetic lens design provides an instrument suitable for both LEEM and PEEM. To do LEEM a sector field produced by magnetic lenses uses the Lorentz force to separate an incident electron beam formed in an illumination column from the reflected or backscattered beam. Thus, the optical axes of both the incident beam and the backscattered electrons are perpendicular to the sample surface. The sector angles are either 120$^\circ$ or 180$^\circ$. TL, FL and IL are the transfer, field and intermediate lens of the imaging column. finally projective lens magnify the image onto the detector. To do PEEM, the illumination column is not used and a photon beam is incident directly on the sample surface. The imaging column is identical for both LEEM and PEEM. A schematic is shown in Fig.~\ref{fig:Instruments}(a). In PEEM, the high electron kinetic energy is particularly useful for high magnification, providing that the light source uses micro-focusing optics and that the sample can withstand high photon brilliance. The electrons are close to the optical axis, making aberration correction and thus ultimate spatial resolution possible. However, when energy analysis is also required, the electrons undergo a strong retarding voltage before entering the analyzer with a significant loss in transmission and space-charge can be a limitation. The best spectroscopic resolution is 0.25-0.30 eV. To perform energy-filtered PEEM with this set-up, a hemispherical energy analyzer is added just before the projective lenses, a set-up called a spectroscopic LEEM.\cite{Locatelli2008a} Figure.~\ref{fig:Instruments}(c) shows a typical LEEM-PEEM instrument with magnetic lenses. 
\subsubsection{Electrostatic lenses}
The electrostatic lens design (Fig.~\ref{fig:Instruments}(b)) can only be used for PEEM since a sector field is impossible to implement. \cite{Escher2005a} However, the electrons in the PEEM  imaging column have only moderate kinetic energy meaning a much higher phase space entering the energy analyzer, and thus a much higher transmission. The near ground potential of the sample is a considerable advantage if one needs to apply potentials to the sample, for example to switch the ferroelectric polarization. Furthermore, the use of a double hemispherical analyzer allows low pass energies and hence high spectroscopic resolution without degrading the spatial resolution. The risk of radiation damage from intense photon beams is reduced and space charge is less of a limitation because it is possible to use lower incident flux. Figure.~\ref{fig:Instruments}(d) shows a PEEM with a double hemispherical analyser and electrostatic lenses. Positions 1, 2 and 3 show the detector positions for direct imaging without energy filtering, small area X-ray photoelectron spectroscopy and spectroscopic imaging. The sample is at position 4 and 5 indicates the double hemispherical analyzer.

\subsection{PEEM}
When a photon of energy h$\nu$ is incident on a sample, all of the electrons whose binding energy is smaller than h$\nu$ can undergo a transition from their initial state. For an electron to be photoemitted from the material and detected, it must overcome the work function, $\Phi$, the potential barrier to be overcome to extract the photoelectron.

\begin{figure}[ht]
\includegraphics[width=6.5cm,clip,bb=0 0 312 538]{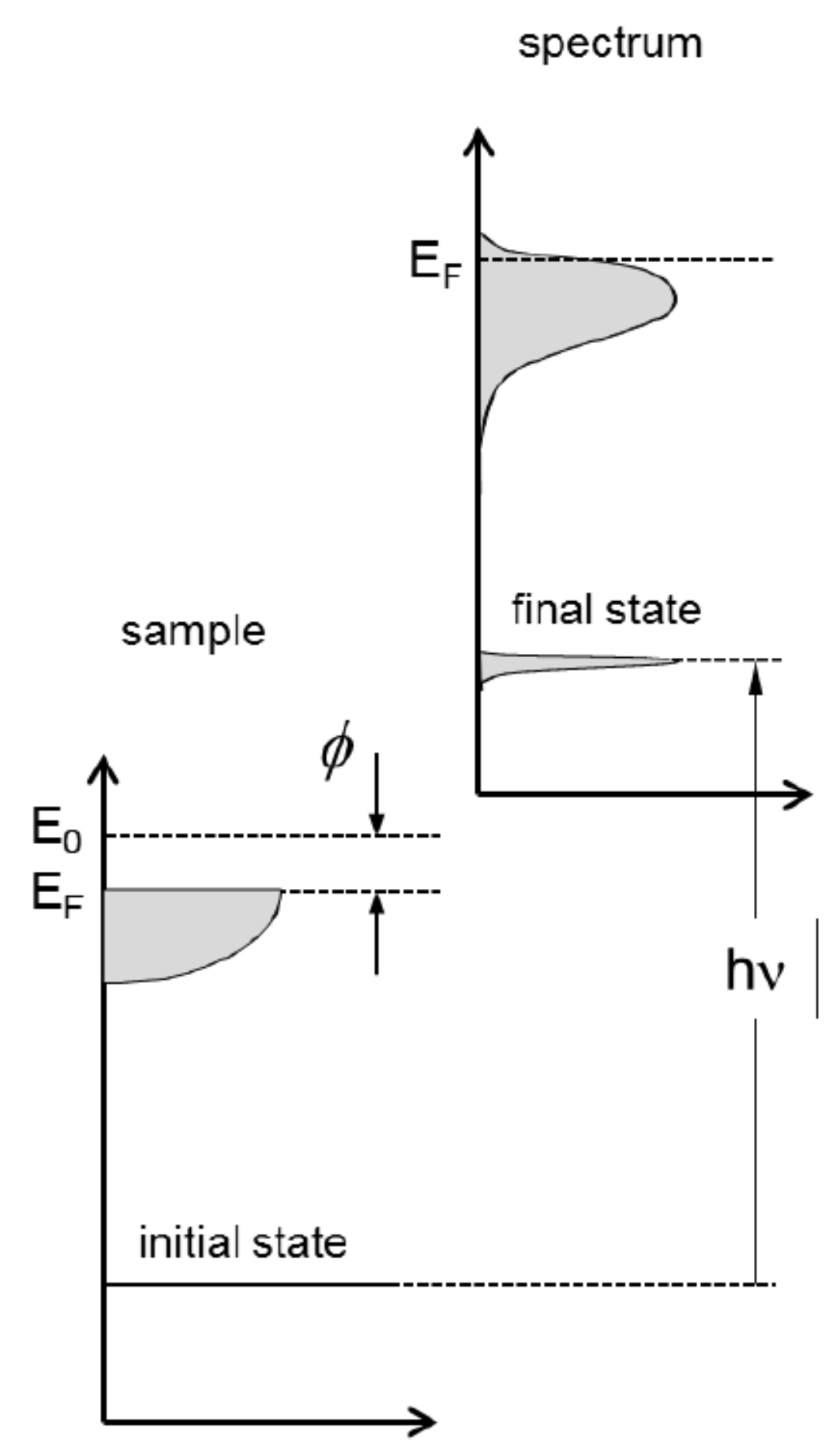}
\caption{(color online) Schematic energy levels in a photoemission experiment. The initial state (left) is excited by a photon h$\nu$, giving the final state as measured by the spectrometer (right). The work function, $\phi$, is the difference between the local vacuum level and the sample Fermi energy, $\mathrm{E_0}-\mathrm{E_F}$.}
\label{fig:PES}
\end{figure}

The photoemission process is represented schematically in Fig.~\ref{fig:PES}. To know the electronic and chemical structure of a material one has to deduce the initial state of the electronc (left hand side of Fig.~\ref{fig:PES}) from the measured photoemission spectrum (right hand side of Fig.~\ref{fig:PES}). In the solid, electrons occupy discrete, localized states (core levels) or more delocalized states (valence band, and for metals, the conduction band). The binding energy is specific to the emitting atom and is a sensitive signature of the local chemical and electronic environment. The valence electrons are responsible for the electronic structure and reflect the insulating, semiconducting or metallic nature of the material. It is also the valence electrons, and for conducting samples, the states at the Fermi level, which determine many key properties for technological applications, such as electron transport, magnetization, spin polarization, or electron-electron correlation. Thus, from a photoelectron spectrum, knowledge of the work function $\phi$ ($\phi=\mathrm{E}_F - \mathrm{E}_0$, where $\mathrm{E}_F$ is the Fermi level and $\mathrm{E}_0$ the vacuum level)and the photon energy allows deduction of the initial states of the analyzed electrons. \cite{Hufner2005}

Since the beginning of X-ray Photoelectron Spectroscopy (XPS) with Siegbahn's Nobel Prize winning work in the late 1950's,\cite{Nordling1957} XPS has evolved into a standard technique for acquiring chemical and elemental information of surfaces \cite{Briggs2003, Hufner2005}. Although first attempts in acquiring lateral information with XPS started more than 20 years ago, these techniques have not, as yet, achieved widespread use. Laboratory imaging XPS are slow and spatial resolution is typically $\simeq 10-20 \mu$m for reasonable acquisition times. Photoemission microscopes [10,11] improve this figure by some two orders of magnitude in the laboratory and three orders of magnitude using synchrotron radiation. A typical field of view (FoV), not to be confused with the lateral resolution, is several tens of microns. However, a direct consequence of the high lateral resolution is a considerably reduced overall transmission of the electron optical system. Thus, the use of these instruments for laboratory experiments may be limited to low energy applications in the Ultra-violet Photoelectron Spectroscopy (UPS) range. This disadvantage can be overcome by using high brilliance undulator beamlines at the synchrotron for imaging XPS experiments.

\begin{figure*}
\includegraphics[width=16cm,clip,bb=0 0 682 262]{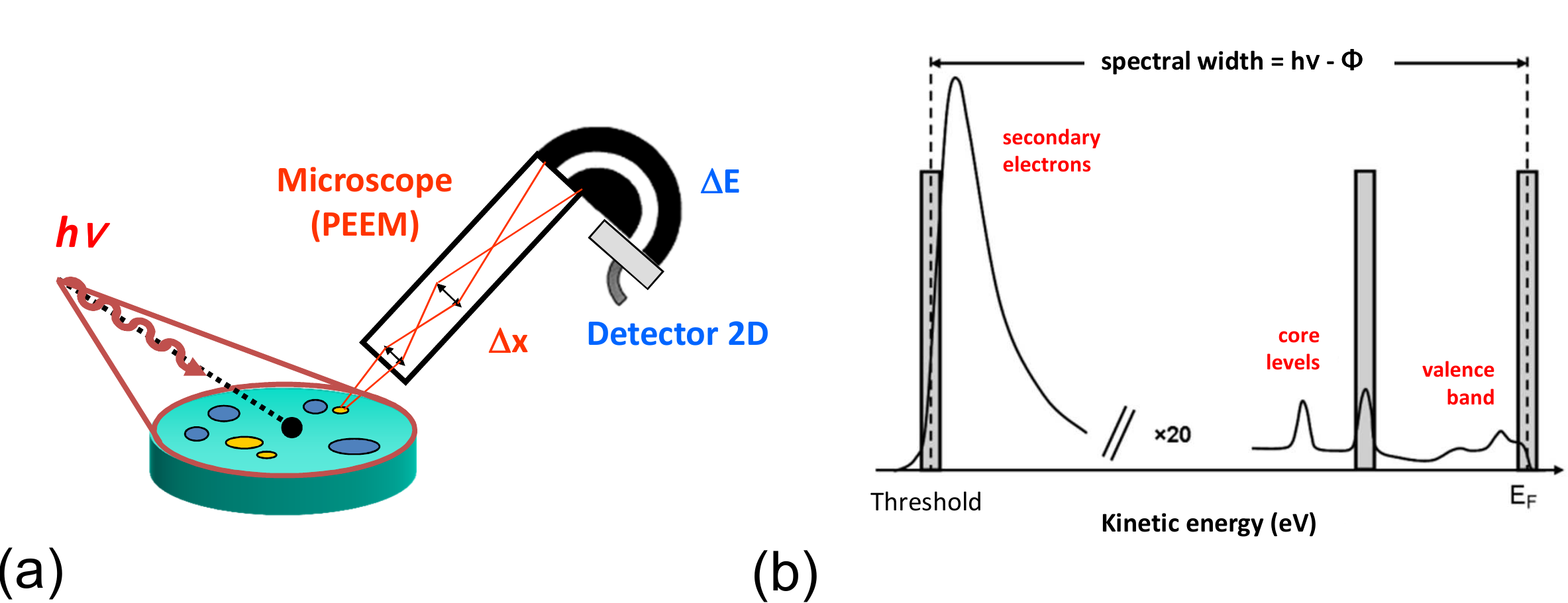}
\caption{(color online) (a) Schematic of a spectromicroscopy experiment. Electron optics conserve the provenance of the electrons in the sample field of view. The energy analyzer filters the kinetic energy, giving a spectroscopic image of the photoemission on the 2D detector. (b) Typical photoemission spectrum. The shaded rectangles highlight the high intensity photoemission threshold at low kinetic energy, the core level region and the valence band.  In PEEM the kinetic energy is referenced with respect to the Fermi level of the sample holder.}
\label{fig:SchematicPEEM}
\end{figure*}

PEEM can be performed just with electron optics and a 2D detector, recording the secondary electron current. Early PEEM work has been carried out using such a system. Today, still many PEEM experiments focussing on absorption spectroscopy measure the electron yield and do not require an energy analyser. By adding an energy filter to the electron optics, the image of the sample surface in the FoV is obtained at a well-defined electron kinetic energy. The same 2D detection system is used to provide an image of the microscopic field of view. Such a spectromicroscopy set-up is shown schematically in Fig.~\ref{fig:SchematicPEEM}(a). The energy filter allows selected regions of the photoemission spectrum to be imaged. Energy regions of interest such as the photoemission threshold, core level and valence band regions are highlighted by the shaded rectangles in Fig.~\ref{fig:SchematicPEEM}(b). There is a slight variation in energy, called non-isochromaticity, along the vertical axis, corresponding to the dispersive plane of the analyzer, but this is usually small, typically a fraction of an electron-volt, and is accounted for in data analysis.

In energy-filtered operation, image stacks are recorded by scanning the sample voltage with fixed analyzer pass energy and automatic refocusing of the objective lens. The three-dimensional data stack, $I(x,y,E)$, therefore contains in each image-pixel microscopic and spectroscopic information, and can be analyzed off-line by standard data reduction techniques, e.g. removal of photoemission background or principal component analysis to reduce the noise in core level images. \cite{Walton2005} The data sets are easily exploited by choosing any region of interest and extracting the corresponding $I(E)$ curve.

Using the electrostatic lens design, the sample is referenced to ground, thus with knowledge of the analyser work function, the onset of the photoemission spectrum at low energy is precisely the work function of the sample surface. \cite{Bailly2008, Zagonel2009} In the magnetic lens design, the energy is referenced to the electron source work function.

\subsection{k-PEEM}
The standard method for measuring the dispersion relations of the band structure is Angular Resolved PhotoElectron Spectroscopy (ARPES). In ARPES, the energy-resolved photoelectron intensity distribution is usually measured as a function of angle along a high symmetry direction of the crystal. Conservation of the component of the electron wave vector parallel to the sample surface allows one to map out the dispersion relations as a function of the initial state wave-vector. In practical units the wave vector is given by  $\mathrm{k}_{parallel} = 0.512\sqrt{\mathrm{E}_{K}}sin\theta$, where $\mathrm{E}_{K}$ is the photoelectron kinetic energy and $\theta$ the photoelectron emission angle with respect to the surface normal. In order to map the dispersion relations over the full Brillouin zone the sample must be rotated through $180^{\circ}$ resulting in long acquisition times.
In PEEM, the high extractor field collects the photoelectron emission over a wide angular range. By using a suitable transfer lens the diffraction plane (close to the back-focal plane) of the PEEM can be imaged. Imaging the diffraction plane in PEEM produces an angular intensity map for all azimuths simultaneously. Energy-filtering transforms the angular distribution into a map of photoelectron intensity as a function of wave-vector parallel to the surface, i.e. a horizontal, constant energy cut in reciprocal space. This technique is known as k-resolved PhotoEmission Electron Microscopy (k-PEEM). For example, the Fermi surface can be acquired in a single-shot experiment. \cite{Kroemker2008} For a typical PEEM settings, the reciprocal space dimension of the image is $\pm 2.5 \text{\AA}^{-1}$. Thus, this technique allows visualization of the full 2D structure of the electron dispersion relations from micron sized regions or domains, see, for example, a recent study of few-layer graphene on SiC.\cite{Mathieu2011b} An important variant of reciprocal space imaging is the use of an aperture in an intermediate image plane which allows one to perform direct $\mu$-ARPES on micron scale regions.\cite{Mentes2012} As in real space PEEM described above, the data sets are also three dimensional, $I(k_x,k_y, E)$.

\subsection{MEM}
In MEM the incident electrons are retarded to a few volts, called the start voltage (SV) by the high sample bias. At very low SV, electrons are reflected by the potential above the surface (MEM), whereas at higher values they penetrate the sample and are backscattered (LEEM). Mirror electron microscopy (MEM) using a low energy electron microscope, allows non-contact, full-field imaging of the surface topography and potential with 12-15 nm spatial resolution. The transition from the reflection of the electrons to the back scattering regime, the MEM-LEEM transition, images the electrostatic potential above the surface. MEM contrast is therefore related to work function differences and it is an ideal tool to probe, for example, surface charge differences in domains with a polarization component perpendicular to the surface, pointing either inwards ($\text{P}^{-}$) or outwards ($\text{P}^{+}$).

At the surface, an in-plane electric field will be created at the boundary between oppositely polarized domains due to the fixed charge. Depending on the width of the space charge region around the domain wall, this field can attain values significant with respect to the extractor field, deviating the reflected electrons in a direction parallel to the surface. The deviation due to the built-in field at a pn junction has already been used in emission microscopy. \cite{Nepijko2002, Nepijko2003} Thus MEM can be used to study the full surface electrical topography.

The first demonstration of MEM to probe the surface potential of differently polarized ferroelectric domains was reported by Spivak et al. using an extremely simple electron lens and single crystal $\text{BaTiO}_3$. \cite{Spivak1959} Cherifi et al. reported polarization induced contrast in thin, epitaxial $\text{BiFeO}_3$ (BFO) films with PFM written domains. \cite{Cherifi2010a} One of the main challenges is the quantification of the surface charge from the measurement of the surface potential, since the latter decreases away from the surface.

\subsection{LEEM}
LEEM images surfaces using elastically backscattered low energy electrons. For single crystal systems both real and reciprocal space information is available. The low kinetic energy makes LEEM, like photoemission, intrinsically sensitive to the surface and near surface region. LEEM gives information not only on the surface morphology, but it is also sensitive to the electronic and crystal structure. In-situ crystal growth or even surface diffusion processes can be monitored. Furthermore, by scanning the start voltage, the empty states in the conduction band are probed. This has been used to explore quantum oscillations in the back-scattered electron signal for 2D systems. The results can be directly interpreted in terms of the conduction band structure, for example on few-layer graphene.\cite{Hibino2008} Finally, if the focal or diffraction plane is imaged in LEEM, then one obtains a $\mu$-LEED image of the sample surface. \cite{Altman2010} The datasets are in the form of 3D stacks: $I(x,y, SV)$ for LEEM and MEM; $I(k_x,k_y, SV)$ for $\mu$-LEED.

\section{APPLICATIONS}
In this section we present five studies of FE systems using full field electron spectromicroscopy. The first study was done using a NanoESCA installed at the TEMPO beamline at the SOLEIL synchrotron (Saint Aubin, France). The screening study of PZT and the band structure of BTO were performed at the Nanospectroscopy beamline at the Elettra synchrotron (Trieste, Italy) using a spectroscopic LEEM. The photogenerated charge screening study was carried out using a laboratory instrument at the CEA Saclay (France). The study of the thickness dependence of the polarization in $\text{BiFeO}_3$ used a NanoESCA instrument at the Omicron Nanotechnology premises in Taunusstein(Germany) and the CEA Saclay laboratory instrument.

\subsection{Polar domain contrast}
 
\begin{figure*} [ht]
\includegraphics[width=16cm,clip,bb=0 0 737 186]{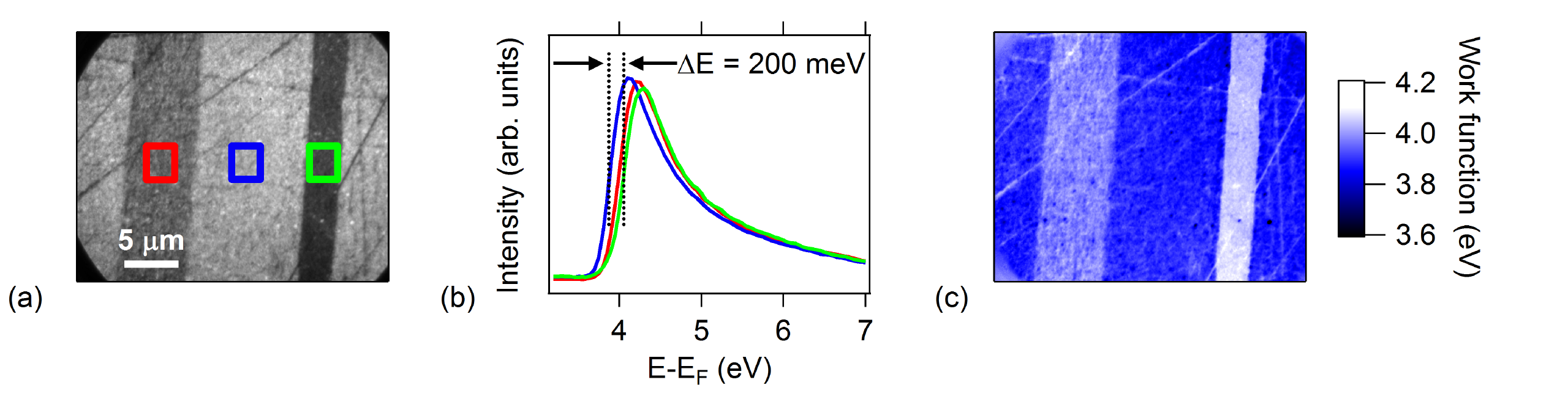}
\caption{(color online) (a) Energy-filtered, threshold image of a $\text{BaTiO}_3$ (BTO)(001) single crystal showing three main intensity levels, $\text{E}-\text{E}_{\text{F}} = 3.85$ eV. (b) Threshold spectra as extracted from the (solid) rectangles in (a). (c) Photoemission threshold map obtained using a pixel-by-pixel fit to the threshold spectra.
}
\label{fig:XPEEM_BTO}
\end{figure*}

The first example shows an unpublished example of how energy-filtered PEEM can be used to identify differently polarized domains at the surface of a ferroelectric. The sample was a $\text{BaTiO}_3$ (001) single crystal. Before introduction into the vacuum system the sample underwent a short exposure ($\sim\!$10 minutes) in air at room temperature to a UV lamp. The UV radiation creates ozone which reacts with the sample surface cleaning it of most organic contamination. An insulator charges during photoemission therefore the sample was then annealed under vacuum at 700\degC producing oxygen vacancies to provide sufficient conductivity whilst maintaining the ferroelectric phase. This is also sufficient to remove any residual surface carbon. \cite{Chen2010} 
The left hand image of Fig.~\ref{fig:XPEEM_BTO} is taken at a kinetic energy of 3.85 eV, measured with respect to the metallic sample holder, with a 34 $\mu$m FoV and 95 eV photon energy using synchrotron radiation.  There are three main levels of intensity at a given energy. At higher kinetic energy there is contrast inversion (not shown), indicating spatially resolved work function variations. Scanning the full threshold spectrum gives a 3D dataset. The threshold spectra for the regions of interest indicated by the rectangles are shown in Fig.~\ref{fig:XPEEM_BTO}(b). They show the characteristic, sharp rising edge of the photoemission spectrum and the broader, secondary electron peak. Three distinct photoemission thresholds are observed. Performing a pixel by pixel analysis, fitting the shape of each threshold spectrum independently gives the threshold map shown in Fig.~\ref{fig:XPEEM_BTO}(c). The values correspond to domains (in order of decreasing threshold) with polarization pointing out from the surface into vacuum, along the surface and into the surface. Thus, energy-filtered PEEM can accurately map the distribution of polar domains at the surface.

Domain recognition by PEEM has recently been used to demonstrate the existence of polar domains in a BTO single crystal well above the the bulk Curie temperature and has been linked to a stabilization of the tetragonal distortion by an ionic surface relaxation.\cite{Hofer2012} 

\subsection{Adsorbate screening}

\begin{figure}[ht]
\includegraphics[width=7.5cm,clip,bb=0 0 225 313]{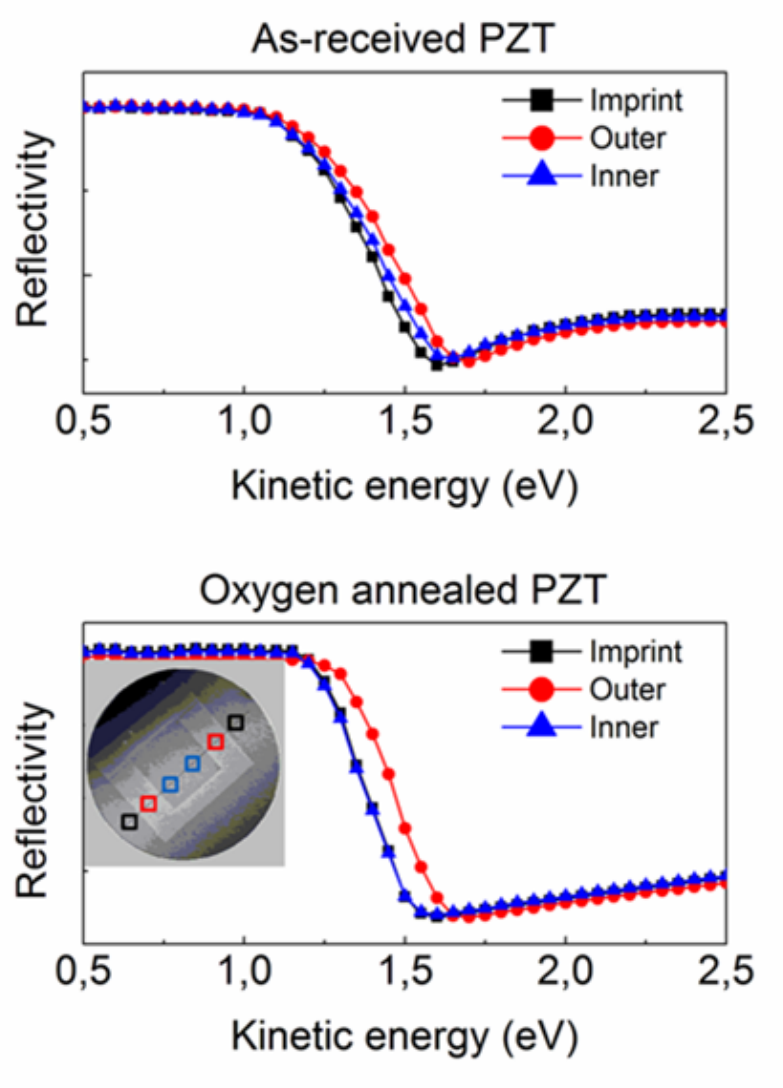}
\caption{(color online) (top) electron reflectivity curves from PFM written $\text{P}^{+}$ and $\text{P}^{-}$ domains in as-received PZT. (bottom) Same after annealing in oxygen. The inset shows the $\text{P}^{+}$ (inner), $\text{P}^{-}$(outer) PFM written squares and the surrounding unwritten "imprint" area. The MEM-LEEM transition is taken as the midpoint in the drop in reflectivity. The FoV is about 15 $\mu$m. \cite{Krug2010} Reprinted with permission from Appl. Phys. Lett. 97, 222903 (2010). Copyright 2010 American Institute of Physics.}
\label{fig:PZT}
\end{figure}

Surface polarization charge in FE materials can be screened by a variety of mechanisms: intrinsic (charge carriers or defects in the bulk), extrinsic (chemical environment or adsorbates),\cite{Fong2006}  domain ordering, or even a combination of the above. For example, chemisorption of $\mathrm{(OH)}^{-}$ and protons can lead to important changes in the electrical boundary conditions \cite{Shin2009} and water film can play an active role in domain switching. \cite{Brugere2011}
Krug et al. studied the effect of surface carbon contamination on the surface charge screening of PFM written micron-scale domains in PZT.\cite{Krug2010} A (001)-oriented $\text{PbZr}_{0.52}\text{Ti}_{0.48}\text{O}_3$ (PZT) layer was grown by high pressure sputtering on a conducting $\text{SrRuO}_3$ (SRO) back electrode which itself was deposited on $\text{SrTiO}_3$(001) single crystal substrate. The PZT stoichiometry was chosen near the morphotropic phase boundary to assure a maximum  piezoelectric response.\cite{Contreras2005}  The layer thickness of 30 nm was below the threshold for strain relaxation, ensuring a low defect density and a high polarization normal to the sample plane. The outer square of (10x10) $\mu$m$^2$ was written by a positive bias of +5 V creating a negative image charge below the surface and thus a $\text{P}^{-}$ state (inward directed polarization). The polarity of an inner (5x5) $\mu$m$^2$ square was reversed by a -5 V bias ($\text{P}^{+}$). The exterior, i.e., unwritten sample area, exhibited a net positive polarity, probably due to a FE imprint in the film after growth and cooling to room temperature. The PZT layer was covered with a platinum mask with open fields to facilitate the location of the FE domains. In addition, the mask provided energy calibration of the photoemission spectra with respect to the Pt 4f peaks and the Fermi level.  Energy-filtered XPEEM at the C 1s core level was used to check the level of surface contamination. Then, the MEM-LEEM transition was measured before and after cleaning by in-situ annealing in oxygen ($4.8\times10^{-5}$mbar at 400\degC for three hours). The electron reflectivity curves for the $\text{P}^{+}$, $\text{P}^{-}$ and as grown areas are shown in Fig.~\ref{fig:PZT}. Before cleaning in oxygen the contrast in the electron reflectivity is low, consistent with the high surface carbon concentration, measured by XPEEM, which provides partial screening of the surface charge. The reduced surface potential difference is ascribed to carbon contamination, screening the contrast expected from written domains with opposite polarization.  After oxygen annealing, clear contrast appears, reflecting a surface potential difference between the $\text{P}^{+}$ and $\text{P}^{-}$ domains. As can be seen from Fig.~\ref{fig:PZT}, the MEM-LEEM transition happens at lower start voltage for the $\text{P}^{+}$ (inner) than for the $\text{P}^{-}$ (outer) sample area, reflecting the fact that the surface potential is higher for the $\text{P}^{+}$ polarity than for $\text{P}^{-}$. The increased contrast on the clean surface is supported by DFT calculations of a nine-layer PbO/$\mathrm{TiO}_2$ stack (without adsorbates) in an external electric field. The stabilized out-of-plane polarization leads to a positive (negative) energy shift of the surface band structure for $\text{P}^{+}$ ($\text{P}^{-}$). Assuming constant electron affinity, the MEM-LEEM transition should then be shifted to lower (higher) start voltages, in qualitative agreement with experiment.\cite{Krug2010}

\subsection{Screening by photogenerated charge carriers}
In this example, the screening by mobile charge carriers was studied. \cite{Wang2012} A BTO(001) single crystal was ozone cleaned before annealing at 700\degC for 3 hours at a base pressure of $1\times10^{-9}$ mbar to remove residual contamination such as $\text{H}_2\text{O}$ or $\text{CO}_{2}$ and to create sufficient oxygen vacancies to avoid charging during measurements. The sample was transferred under vacuum for the MEM-LEEM analysis. A 254 nm UV lamp illuminated the surface with 20$\mathrm{mW/cm^{2}}$ power flux to create electron-hole pairs.

\begin{figure}
\includegraphics[width=8.5cm,clip,bb=0 0 599 383]{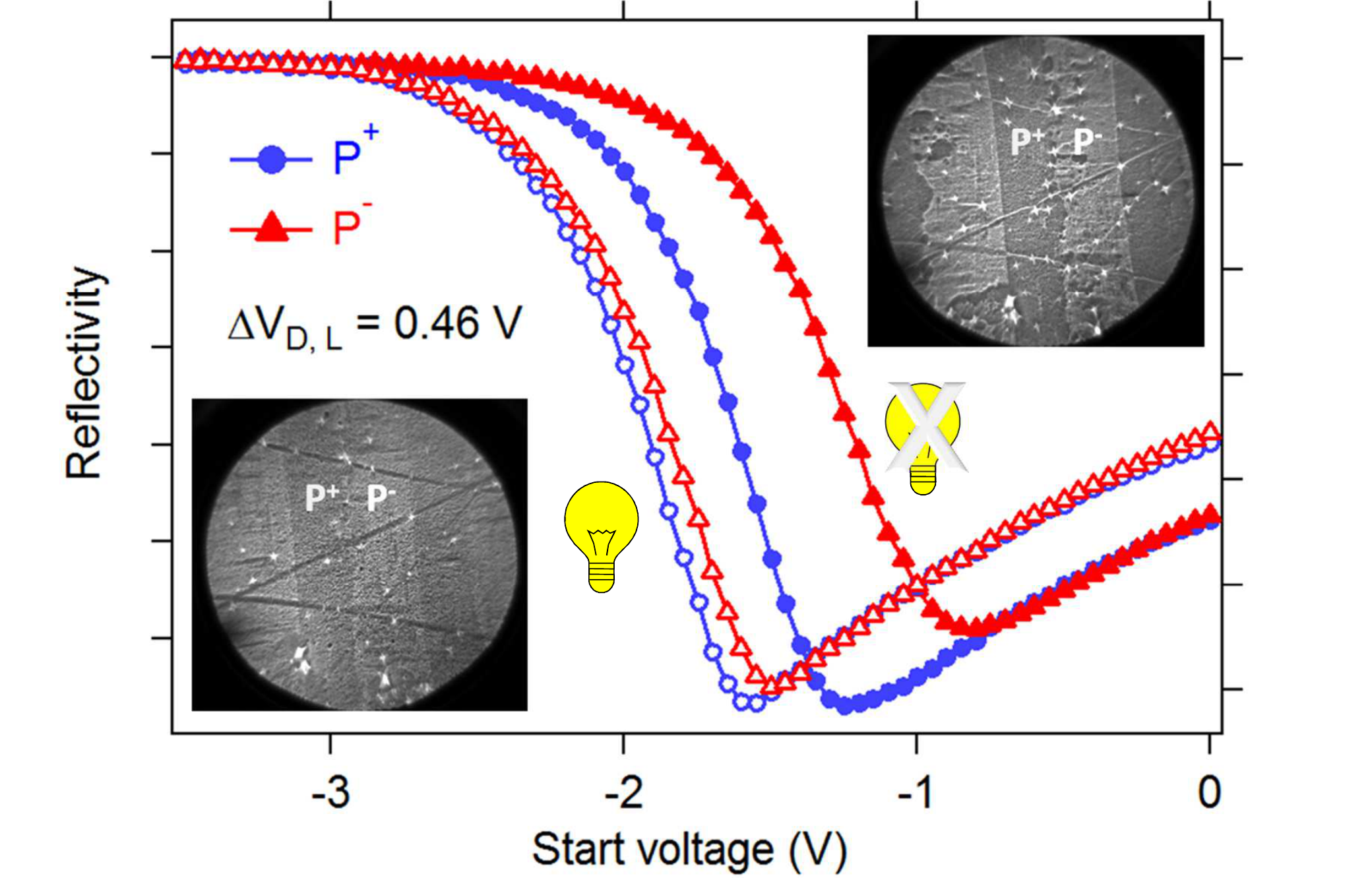}
\caption{(color online) Reflectivity curves before (solid symbols) and after (open symbols) exposure of a BTO(001) single crystal to UV light. Insets: MEM image before (upper right) and after (bottom left) exposure to UV light. \cite{Wang2012} Reprinted with permission from Appl. Phys. Lett. 101, 092902 (2012). Copyright 2012 American Institute of Physics.}
\label{fig:BTO_UV}
\end{figure}

The right hand inset of Fig.~\ref{fig:BTO_UV} is a MEM image from the clean sample at a start voltage of -1.8V in a 40 $\mu$m FoV. Two intensity levels are observed corresponding to different electrostatic potentials above the surface. The intensity contrast inverses as the start voltage is increased. We identify the two regions as $\text{P}^{+}$ and $\text{P}^{-}$ polarized domains (see below). The electron reflectivity of the two domains as a function of start voltage, measured with respect to the reflected intensity at -5 V, is plotted in the main part of Fig.~\ref{fig:BTO_UV} (full symbols). The MEM-LEEM transition is the midpoint of the decrease in the reflectivity curve. For the clean surface, the MEM-LEEM transition shift is 400 mV. Such a large value is attributed to opposite polarizations perpendicular to the surface, pointing either outwards ($\text{P}^{+}$) or inwards ($\text{P}^{-}$). It cannot be due to domains with different magnitudes oriented in the same direction because they do not screen the depolarizing field. A SSPM/PFM study by Shao et al. reported a 150 - 200 mV shift between $\text{c}^{+}$ and $\text{c}^{-}$ domains; however, their data were acquired in air and the potential contrast was probably attenuated by adsorbates.\cite{Shao2006}  After a few tens of seconds of exposure to UV light, the shift in the MEM-LEEM transition between the two domains is strongly reduced, reaching a steady state value of 90 mV after 200 s.

The reflectivity curves after UV exposure are plotted as open symbols in Fig.~\ref{fig:BTO_UV}. The photoemission intensity due to the UV illumination is also detected, however, it was more than three orders of magnitude lower than that of the MEM-LEEM signal and can be neglected. As can be seen in the left hand inset of Fig.~\ref{fig:BTO_UV}, the domain structure is unchanged by the UV illumination but the contrast is weakened. When the UV light is switched off, the contrast reverts to the original value, although at a slower rate. The experiment is repeatable without altering the domain structure. The average value of the MEM-LEEM transition shifts by $\Delta \text{V}_{\text{off,on}} =$ 0.46 V to lower start voltage when illumination is switched on. UV illumination creates not only electron-hole pairs in the material but also induces photoemission from the material. The latter gives rise to a separate effect with respect to the surface charge screening and the photoemission positively charges the surface. The absolute MEM-LEEM transition shift of each domain is therefore consistent with an equal 155 mV change in $\text{P}^{+}$ ($\text{P}^{-}$) surface potential by electron (hole) screening. The shift in the MEM-LEEM transition measures potential change perpendicular to the surface, therefore, in the case of in-plane polarization only a photoemission induced shift to lower start voltage would be expected with no offset due to screening by electron-hole pairs.

\begin{figure}
\includegraphics[width=8.5cm,clip,bb=0 0 417 339]{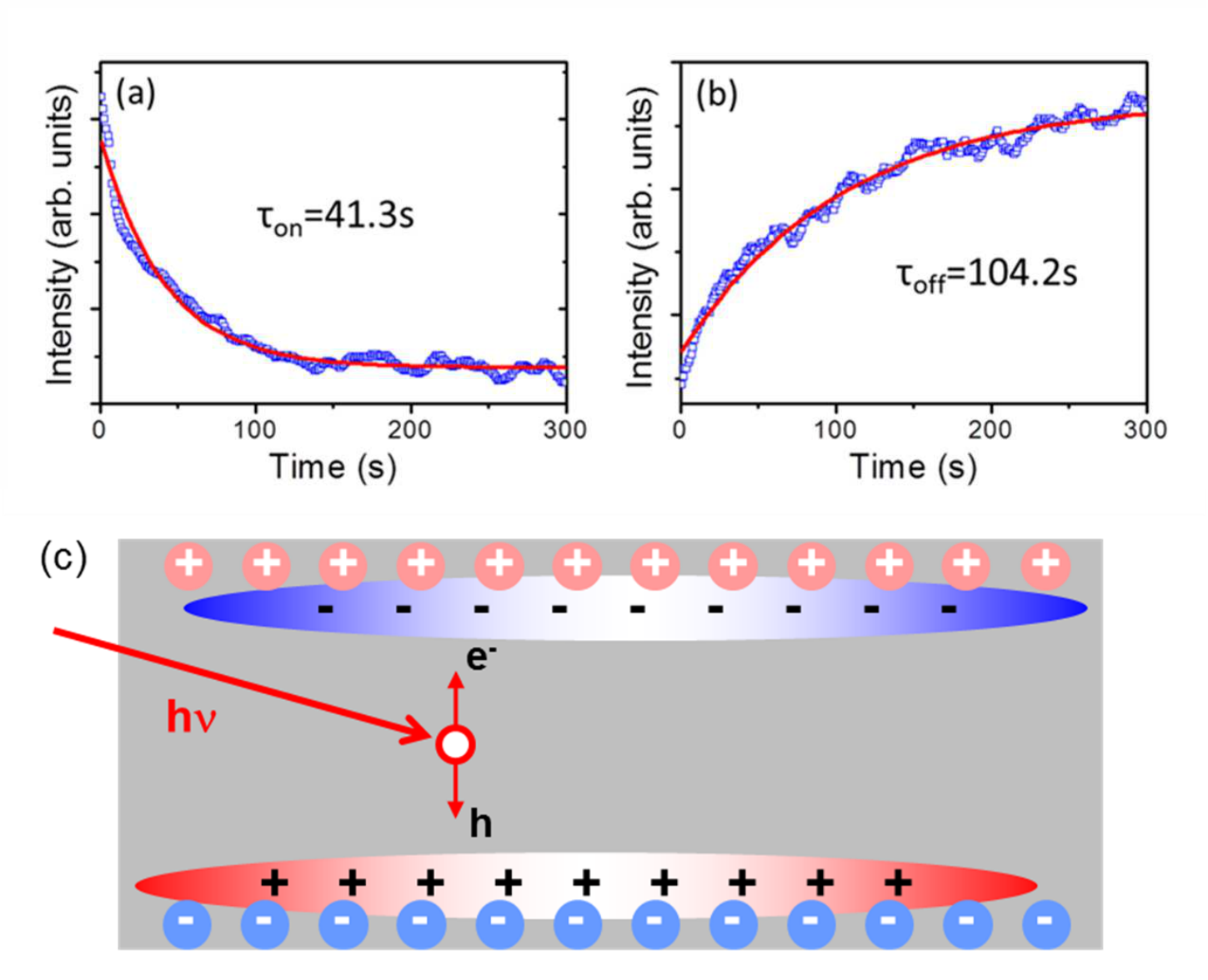}
\caption{(color online) (a) Time dependence of intensity contrast due to screening by UV-generated electron-hole pairs (b) Time dependence of intensity contrast after UV light is switched off. (c) Schematic showing the drift of photogenerated charge carriers in an out of plane polarized ferroelectric slab. \cite{Wang2012} Reprinted with permission from Appl. Phys. Lett. 101, 092902 (2012). Copyright 2012 American Institute of Physics.}
\label{fig:UV_screening}
\end{figure}

To study the surface charge dynamics, we recorded fixed start voltage (-1.8 V) MEM images at 1 frame per second under UV illumination. The time dependence of the domain intensity contrast, defined by $\Delta\text{I} = \text{I}(\text{P}^{+})-\text{I}(\text{P}^{-})$, when the UV light is switched on and off is plotted in Fig.~\ref{fig:UV_screening}(a) and (b). Fitting $\Delta$I using a simple rate equation, $\Delta\mathrm{I} = \mathrm{A} + \mathrm{B}\mathrm{exp}(-t/\tau)$, gives time constants $\tau_{\text{on}} = 41.3$ s for light-on and $\tau_{\text{off}} = 104.2$ s for light-off processes. These values are several times larger than those found by Shao et al.\cite{Shao2006}  with a lower power UV lamp. However, the BTO has been vacuum annealed at 700\degC creating oxygen vacancies which can act as charge traps \cite{Papageorgiou2010}  increasing the time constant. UV illumination generates electron-hole pairs near the surface because of the large photo-excitation cross section. Under the internal polarization field, electrons (holes) drift to the $\text{P}^{+}$($\text{P}^{-}$) surface and some are trapped in vacant trap states as illustrated in Fig.~\ref{fig:UV_screening}(c). Thus, a space-charge field opposing the polarization field increases until equal to the polarization field or until the rate of electron-hole pair generation equals that of recombination. When the light is switched off, the trapped carriers are thermally activated and diffuse along the concentration gradient to eventually recombine. \cite{Shao2006}  The spatial redistribution and retrapping of photocarriers give the observed exponential decays. \cite{Wang2012}

This analysis assumes that the screening by the photogenerated charge only results in a rigid shift of the reflectivity curve and linearity between the MEM-LEEM transition shift and the intensity contrast. The former is clearly a good assumption, while the latter is an approximation. A more refined analysis requires knowledge of the relationship between the transition shift and intensity difference at constant SV or, ideally, the acquisition of reflectivity curves during the screening process. 

\subsection{Polarization of ultra-thin films}
A major issue for prospective nanoscale, strain-engineered ferroelectric applications is the decrease of the remanent polarization of ultra-thin films. Ferroelectric capacitors for example may exhibit a critical thickness.\cite{Kim2005, Petraru2008}  Lichtensteiger et al. \cite{Lichtensteiger2005} have shown that the decrease in remanent polarization of $\text{PbTiO}_3$ (PTO) thin films between 20 and 2.4 nm on Nb-doped STO substrates follows that of the tetragonality (ratio of the out-of-plane to in-plane lattice parameter c/a). On $\text{La}_{0.67}\text{Sr}_{0.33}\text{TiO}_3$ (LSMO) PTO polydomains were formed below 10 nm with high tetragonality.\cite{Lichtensteiger2007}  The formation of a polydomain state has been suggested for SRO/PZT/SRO capacitors with $\text{Pb(Zr,Ti)O}_3$ thicknesses below 15 nm.\cite{Nagarajan2006}  Pertsev and Kohlstedt showed the importance of misfit strain for the critical thickness of the monodomain-polydomain stability for PTO and BTO.\cite{Pertsev2007}  Direct electrical measurements of the polarization-field (P(E)) loop in ultrathin ferroelectric films are a challenge because of leakage current for thicknesses below a few tens of nm.\cite{Bea2006}  They become impossible in the tunneling regime for ultra-thin films (5 nm or less) which, furthermore, is of the same order as the critical thickness estimated from Landau-Ginzburg-Devonshire(LGD) elastic theory for polarization stability.\cite{Bratkovsky2000, Maksymovych2012}  

\begin{figure}
\includegraphics[width=8.5cm,clip,bb=0 0 739 632]{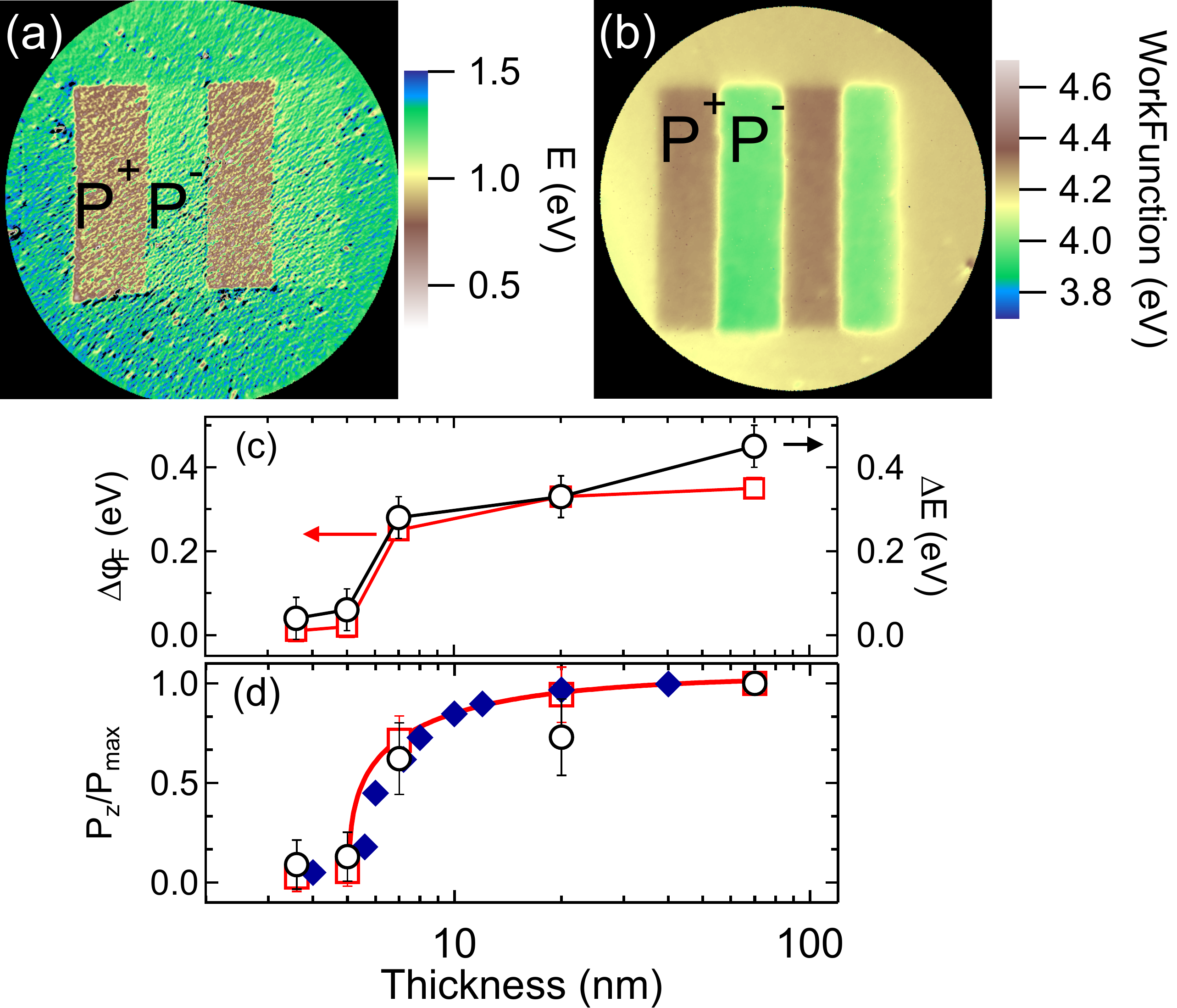}
\caption{(color online) (a) MEM-LEEM transition map obtained by a pixel by pixel analysis of the 3D dataset
(b) PEEM threshold spectra from PFM-polarized $\text{P}^{+}$ and $\text{P}^{-}$ domains in epitaxial BFO. 
(c) Thickness dependence of the MEM-LEEM transition (red squares) and the photoemission threshold (black circles) contrasts as measured by MEM-LEEM and PEEM, respectively. (d) P$_\mathrm{z}$/P$_\mathrm{max}$ calculated from PEEM (red squares) and MEM-LEEM (black circles). Red curve is fit to PEEM/LEEM data with h$_{\mathrm{eff}}$ = 5.6 nm. Blue diamonds are P$_\mathrm{z}$/P$_\mathrm{max}$ values used for numerical simulations. Reprinted with permission from Rault et al.\cite{Rault2012} Copyright 2012 American Physical Society.
}
\label{fig:BFO}
\end{figure}

Rault et al. \cite{Rault2012}  have studied the polarization thickness dependence of strained $\text{BiFeO}_3$ (BFO) films with constant tetragonality. Films in the range 3.6 to 70 nm were studied. Polarized domains were written using PFM and both the work function and electron reflectivity were measured using PEEM and MEM-LEEM. Contrast in the MEM-LEEM transition and the photoemission threshold is clearly observed in Fig.~\ref{fig:BFO}(a) and (b). 
These are not images but MEM-LEEM transition and photoemission threshold maps. They are obtained by fitting the electron reflectivity or threshold spectrum pixel by pixel across the field of view and the plotting the values on a false color scale. The FoV is 40 and 33 $\mu$m, respectively. The benchmark was a 70 nm film whose polarization was measured by classic P-E loop. The polarization of the thinner films can then be estimated from contrast with respect to those of the 70 nm film.

The $\text{P}^{+}$/$\text{P}^{-}$ domain contrasts fall to zero, as shown in Fig.~\ref{fig:BFO}(c), below a critical thickness of 7-8 nm. This compares well with the value of 5.6 nm obtained from the 3D LGD model once the BFO/bottom electrode interface layer thickness is taken into account. The fall in the film polarization is shown in Fig.~\ref{fig:BFO}(d) together with the fit using the 3D LDG theory and a critical thickness of 5.6 nm. Effective Hamiltonian calculations showed that this rapid decrease is due to the stronger depolarizing field, forcing a phase transition from a single to a more stable stripe domain state, explaining why the tetragonality remains constant at the high value of 1.05 while the average domain polarization drops to zero. However, these stripes are predicted to be only a few nanometres wide and are therefore smaller than the typical lateral resolution available. \cite{Rault2012} Aberration corrected LEEM, discussed below, may be able to image directly such a transition. Nevertheless, this example shows that PEEM-LEEM can be used to estimate polarization in micron scale domains even for films in the tunneling regime.

\subsection{Band structure}
\begin{figure*}
\includegraphics[width=14cm,clip,bb=0 0 482 360]{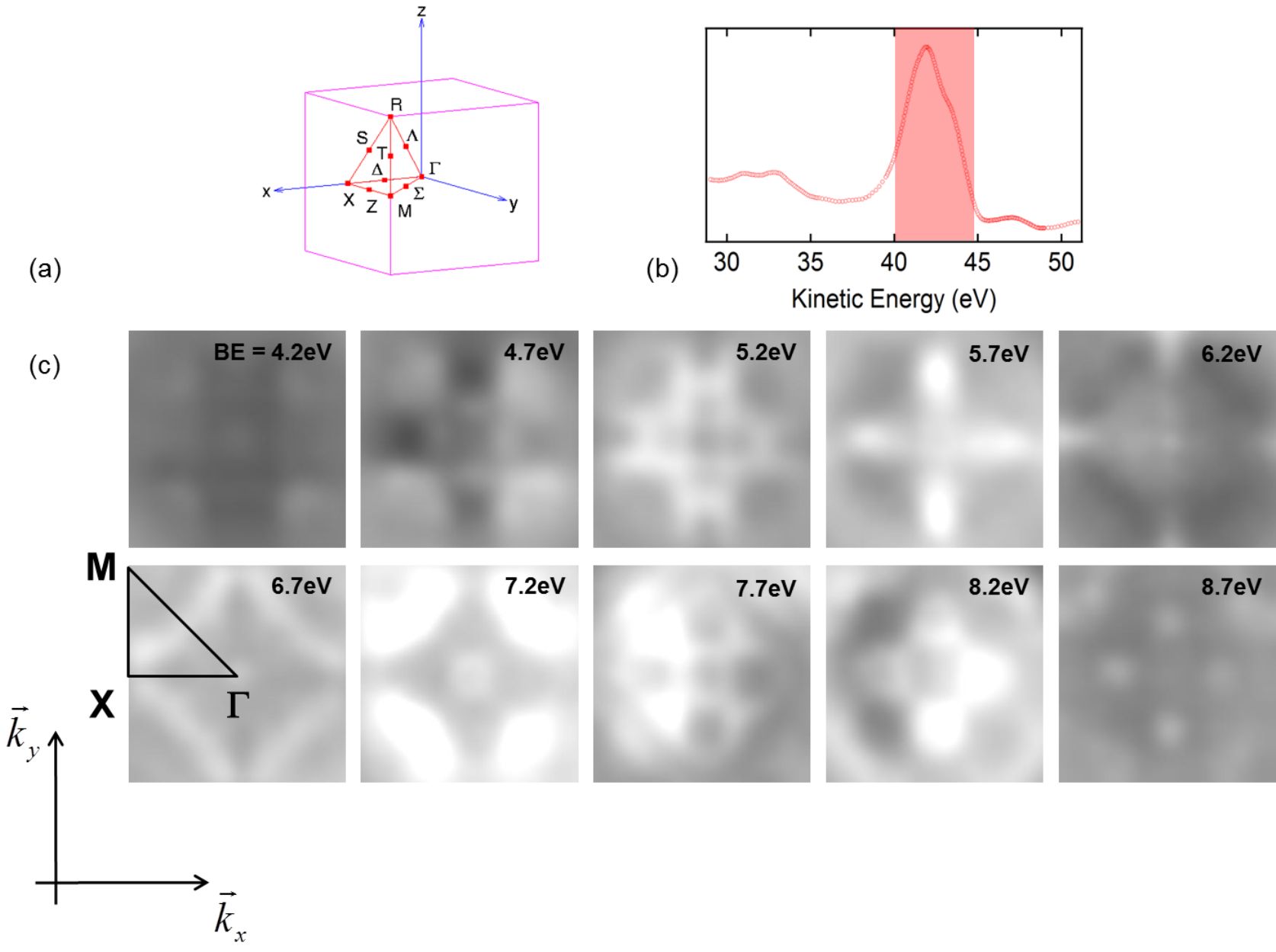}
\caption{(color online) (a) Schematic of the Brillouin zone for BTO(001). (b) Integrated photoemission intensity measured from a single in-plane domain. The sample holder Fermi level is at 48.75 eV. The wave-vector resolved constant energy cuts were acquired in the shaded region (c) k-resolved constant energy cuts in reciprocal space through the valence band (shaded area in (b)) showing the rich band structure.}
\label{fig:BandStructure}
\end{figure*}

Measuring the dispersion relations of the electronic bands in oxides is notoriously difficult even with highly ordered samples because of charging under photoemission. The few examples of angle-resolved photoelectron spectroscopy studies of oxides in the literature concern either metallic oxides, such as niobium doped SrTiO$_\mathrm{3}$\cite{Takizawa2009} or insulating oxides such as SrTiO$_\mathrm{3}$ with a metallic 2D electron gas at the surface.\cite{Meevasana2011,Santander-Syro2011}  The final example illustrates the use of reciprocal space PEEM, k-PEEM, to obtain the full 2D band structure of a single ferroelectric domain. The sample preparation followed the ozone protocol and in-situ vacuum annealing at 700\degC described previously in order to obtain a clean surface with sufficient oxygen vacancies to avoid substantial charging. The instrument was adjusted so that a single in-plane domain filled the field of view. The transfer lens of the optics was then used to obtain an energy filtered image of the back focal plane. The sample azimuth was determined by the $\mu$-LEED mode showing that the domain has a $1\times1$ surface structure. This calibrates the dimensions and orientation of the reciprocal space image.  Then, the undulator beam was switched on for PEEM. By scanning the photoelectron kinetic energy for incident photon energy of 52 eV, constant energy cuts in reciprocal space were obtained from the top to the bottom of the valence band. These are shown as 2nd derivative maps of the intensity as a function of electron wave-vector parallel to the surface in Fig.~\ref{fig:BandStructure}(c) These are preliminary results and a fuller version will be published. 

The integrated intensity (Fig.~\ref{fig:BandStructure}(b)) is typical of BTO as obtained by ultra-violet photoelectron spectroscopy; however, the constant energy cuts in reciprocal space (Fig.~\ref{fig:BandStructure}(c)) reveal the rich band structure of BTO. Between 7.7 and 8.2 eV in particular there is a horizontal mirror plane but no vertical mirror plane in the Brillouin zone. This may be spectroscopic evidence that the imaged domain is indeed in-plane polarized, i.e. the tetragonal distortion is parallel to the surface inducing the asymmetry in the band structure. We note that typical ARPES explores only a single high symmetry direction, for example $\Gamma$X, whereas in k-PEEM the 3D dataset allows simultaneous exploration of the electron bands for all $\mathrm{k}_{parallel}$ wave-vectors. It is therefore possible to explore the details of the electronic band structure, for example wave function symmetry across the Brillouin zone, in a single ferroelectric domain. At present the minimum FoV is about 1-2 $\mu$m, limited by the electron optics.

\section{ADVANTAGES AND LIMITS}
\subsection{Perspectives for ferroelectrics}
Spectromicroscopy is a full field technique, thus no scanning of the sample is necessary. The relatively large fields of view with respect to a typical PFM image will often allow simultaneous visualization of many domains. Dynamical experiments of domain ordering are therefore possible.
Both PEEM and MEM/LEEM are non-contact techniques, avoiding possible unwanted tip-surface interactions of near field methods such as PFM and Conductive Atomic Force Microscopy (CAFM). The spectroscopic capability obtained by energy-filtering in PEEM gives direct access to the initial states of the electrons. Thus a full description of the electronic and chemical structure is possible.
PEEM may be performed with a variety of lights sources: in the laboratory, with lasers or with synchrotron radiation. The use of synchrotron radiation allows tuning of the photon energy: optimization of the photoelectron yield, adjustment of the electron kinetic energy, thus the depth sensitivity. Most undulator beamlines allow variable polarization, thus circularly polarized light can be used to perform X-ray magnetic circular dichroism based PEEM to study ferromagnetic states, whereas linear dichroism allows study of the ferroelectric distortion, for example at the Ti $\text{L}_{2,3}$ edge.\cite{Arenholz2010} 
Finally, the wave-vector resolution afforded in the k-PEEM mode allows immediate visualization of the dispersion relations in the surface plane.

\subsection{Surface preparation}
PEEM and LEEM experiments must be carried out in ultra high vacuum (UHV), mainly to avoid arcing between the sample and the objective lens, which would create high instantaneous currents, potentially damaging to the sample. However, an UHV environment has the advantage that it is possible to study relatively clean surfaces, free of extrinsic adsorbates like water which are inevitably present in air based experiments and which can dramatically change the electrical boundary conditions. Controlled surface preparation is therefore crucial. Traditional surface science techniques such as ion beam sputtering to clean the surface are excluded because of the damage they cause to the surface structure. Careful annealing in $10^{-5} - 10^{-6}$ mbar oxygen environment may be sufficient to clean the sample surface in-situ. \cite{Schrafanek2008}   We have developed a protocol of ex-situ cleaning using a rapid exposure to ozone, immediately prior to introducing the sample into the UHV system. However, without doubt, the very best method would be to assemble the growth (PLD/MBE), characterization (PFM) and PEEM/LEEM analysis tools in the same UHV system, with \textit{in-situ} transfer.
A second important problem is the insulating nature of ferroelectric materials. Some way is necessary to avoid or at least reduce charging either by electrons or as a result of photoemission. In the case of electron charging in LEEM, one solution proposed has been to combine two electron beams of widely different energies. At low energy, the sample is negatively charged whereas at higher energies the number of secondary electrons created is such that the sample charges positively. In principle it is therefore possible to find an equilibrium position of zero charge. Another possibility is to anneal samples to create sufficient oxygen vacancies to make the sample less insulating. Each oxygen vacancy frees two electrons which reduce, in the case of BTO, two Ti$^{4+}$ ions to Ti$^{3+}$. However, care must be taken not to overdope the sample. A recent theoretical study of BTO suggests that beyond a doping limit of $1.36\times10^{21}$cm$^\mathrm{-3}$ a tetragonal to cubic phase transition occurs.\cite{Iwazaki2012} 

\subsection{Lateral resolution and depth sensitivity}
There are several contributions to the limits in spatial resolution of PEEM and LEEM. In PEEM, the most frequently encountered problem is simply the signal level. In many cases it is the transmission of the detector (PEEM optics, energy filter) which reduces the statistics and is often responsible for the practical spatial resolution. \cite{Bailly2009}   This is one of the reasons for preferring more intense synchrotron radiation sources. By reducing the photoelectron kinetic energy the transmission of the PEEM system can also be increased, furthermore the photoionization cross-section can be optimized by adjusting the photon energy. The key point is to increase the ratio of \textit{useful} to \textit{total} photoelectrons entering the PEEM optics. In this respect, successful aberration correction would significantly improve instrumental transmission. \cite{Locatelli2011} In LEEM, counting statistics are rarely a problem because of the high intensity, monochromatic incident electron beam.

The second limitation to the spatial resolution comes from imperfections in the electron lenses, giving rise to chromatic and spherical aberrations. \cite{Escher2010}   High electron energies or smaller fields of view can reduce these, but in the latter case intensity can become a problem. Some attempts at aberration correction have been made, notably for electron optics systems based on magnetic lenses, however, as yet, aberration correction has only been proven in LEEM and routine use remains challenging. \cite{Tromp2010}

The typical lateral resolution in energy-filtered PEEM of core levels is of the order of 100 nm. This improves to 50 nm for measurements of the work function, mainly due to the much higher signal. In LEEM, 20 nm is a routine value. Although generally worse than the spatial resolution obtainable in PFM, individual domains can still easily be imaged and studied. Directly visualizing domain walls, typically a few nm wide, is not possible in PEEM and LEEM, however, the space charge region created at charged domain walls may be resolved. Indeed, the photoemission threshold map in Fig.~\ref{fig:BFO}(b) shows evidence of space charge at the boundary between $\text{P}^{+}$ and $\text{P}^{-}$ domains.

One might think that the high extractor field, typically 100 kV/cm, could switch the polarization in thin films. We have never observed this for a wide variety of ferroelectric materials in both single crystal and thin film forms. The geometry of the sample holder means that all of the sample is at the same potential (close to ground in the case of an electrostatic PEEM), thus the potential difference across a film due to the extractor field is zero.

As in XPS, the moderate photoelectron kinetic energies used in PEEM limit the depth sensitivity to a few nm. PEEM is particularly sensitive to surface chemistry and electronic structure. If one wishes to extend the method to the study of buried, heterogeneous interfaces, for example between an electrode and a ferroelectric, higher energy photons will be necessary. This is extremely demanding for PEEM because of the rapid decrease of the instrumental transmission function with electron energy. Nevertheless, some hard X-ray PEEM has been performed; \cite{Schneider2012} and the use of sensitive, event counting detector systems now make this a real possibility.

\subsection{Beam effects}
High intensity undulator radiation can also have a number of unwanted effects. The incident photon beam results in high rates of electron-hole pair creation. \cite{Wu2011}  This may be useful if one wants to study screening phenomena like in the example given above \cite{Wang2012} or to study the chemistry of catalysis. On the other hand, if longer acquisition times are required for a study of core level binding energies, electron-hole pairs can screen surface charge and thus the internal field, resulting in a shift in the measured binding energies. \cite{Wu2011} This problem has to be assessed for each sample. For example, thin film samples can sometimes be sufficiently conductive to easily evacuate photogenerated charge.

Despite the desirability of high intensity photon beams, one cannot indefinitely increase the photon brilliance in the microscope field of view. Locatelli et al. have shown that beyond $10^{13}$ photons/s, space charge effects occur either at the sample surface or near the first cross-over point of the electron optics, distorting the image. \cite{Locatelli2011}  Fortunately, most synchrotron beamlines are just below this threshold, but when using micro-focusing care must be taken.

Sample charging due to the emission of photoelectrons in PEEM, or due to the injection of electrons from the LEEM can also distort images or cause spectacular spectral shifts, particularly for highly stoichiometric, insulating samples, i.e. with few defects, oxygen vacancies or for single crystals. Again the seriousness of the problem must be assessed in each case. One possibility is to carefully anneal in vacuum single crystal samples as we have done for BTO(001) to create sufficient oxygen vacancies whilst maintaining the ferroelectric phase. Another way to overcome charging is to heat the sample during measurement, providing that this does not modify, for example, the ferroelectric phase being studied.

Finally, direct beam damage must also be considered. The highest intensity  and hard X-ray beamlines can deliver incident beam currents of the order of 1 nA. In high vacuum the energy transferred to the sample can easily give rise to reduction of cation species. Elements such as Pb in PZT or Bi in BFO are particularly susceptible to metallization.

\section{OUTLOOK}
Most synchrotrons now offer PEEM instrumentation; however, access is not always straightforward because of the high demand for beam time. Furthermore, proposals to study ferroelectrics are intrinsically more complex because of the insulating nature of the materials involved and their high sensitivity to intense undulator radiation. The use of laboratory based PEEM, for example with UV lamps or focused He I/II discharge sources may therefore be an interesting alternative. Laboratory based lasers, particularly if frequency doubled and used in the 2 or 3 photon mode may also be an extremely good alternative light source. However, laser based PEEM requires careful attention to average and peak power outputs. In general, high repetition rates (MHz) are mandatory in order to avoid space charge effects in the PEEM optics. \cite{Buckanie2009}  Femtosecond lasers could be used to study, for example, domain switching, where the switching times of the local atomic distortions are on the scale of the nanosecond.

It has been accepted to a large extent that the polarization state gives rise to a rigid shift in the electron bands, similar to the effect of doping in semiconductors. However, there is no reason why this must always be the case. In BTO, the distortion around the $\text{TiO}_6$ octahedra is principally responsible for the soft phonon mode associated with the ferroelectric state. In PZT it is the movement of the Pb atoms which makes the biggest contribution, i.e. the type A cation rather than the B-cation. The work on dynamical charge, notably by Ghosez, \cite{Ghosez1998} emphasizes that the measurable quantity is the change in polarization associated with an atomic distortion. Thus rather than discussing the effective valence state of an ion one should think in terms of dynamical charge tensors.  These can be quite large, and very different from one atom to another and one orbital to another.  A study of the change in the core level binding energies with polarization should therefore yield precious data on dynamical charge tensors.

Reciprocal space PEEM, k-PEEM, could become an extremely powerful tool since much of the research of magnetoelectric coupling is starting to focus on orbital overlaps \cite{Vaz2010} or hybridization \cite{Niranjan2008} at the interface between a magnetic and ferroelectric material. More generally, such a technique could provide valuable experimental data to compare with first principles calculations of, for example, the electronic structure of a ferroelectric domain in contact with a metal electrode. \cite{Stengel2011}

Magnetoelectric coupling between a ferromagnetic and a ferroelectic could be directly imaged by a combination of X-ray absorption and photoemission PEEM. The growing interest in the use of multiferroic materials for spintronics \cite{Allibe2012} means that spin polarized PEEM and LEEM are exciting prospects. \cite{Kronast2010, Suzuki2010} Spin polarized LEEM can probe both in and out of plane magnetization by means of the spin dependent exchange scattering. \cite{Man2003}

A final direction of research which will benefit from PEEM and LEEM techniques is the behavior and the characterization of structures on device scales. Using suitable, customized sample holders, it is possible to wire up for example a single microscopic capacitor and, using the high spatial resolution study its response to current or voltage. \cite{Heron2011} The use of hard X-rays could provide access to the chemistry and electronic structure of  both electrode/ferroelectric interfaces in asymmetric, microscopic capacitors.\cite{Wiemann2012} Nanoscale control of domain walls in structured ferroelectric films is now leading to new applications in catalysis or photovoltaics.\cite{Yang2010} By the addition of suitable electrodes and contacts, device responses on the microscopic scale could be studied using spectromicroscopy under different electrical conditions.


%
%

%

\begin{acknowledgments}
We thank K. Winkler and B. Kr\"{o}mker (Omicron Nanotechnology) for support of some of the PEEM measurements, D. Martinotti for technical assistance with the LEEM experiments at the CEA-Saclay and E. Jacquet, C. Carr\'{e}t\'{e}ro and H. B\'{e}a for help in sample preparation at CNRS/Thal\`{e}s. We acknowledge Elettra and SOLEIL for provision of synchrotron radiation and the Nanospectroscopy and TEMPO beamline staff for support. This work was supported by the French National Research Agency (ANR) projects Surf-FER, M\'{e}lo\"{i}c, Multidolls and Nomilops. C. M. benefited from a grant of the CEA Nanosciences programme. J. R. is funded by a CEA PhD grant. W. R. acknowledges the Eastern Scholar Professorship at Shanghai Institutions of Higher Education, Shanghai Municipal Education Commission, and support from National Natural Science Foundation of China under Grant No. 11274222.
L. B. thanks the financial support of ARO Grant No. W911NF-12-1-0085, and ONR Grants No. N00014-11-1-0384, N00014-12-1-1034 and
No. N00014-08-1-091. L.B. also acknowledges NSF DMR-1066158 and DMR-0701558, and Department of Energy, Office of Basic Energy Sciences, under contract ER-46612 for discussions with scientists sponsored by these grants.
\end{acknowledgments}

\end{document}